\documentclass{ieeeaccess}
\usepackage{cite}
\usepackage{caption}
\usepackage{amsmath,amssymb,amsfonts}
\usepackage{verbatim}

\usepackage{soul}
\soulregister\cite7
\soulregister\ref7
\soulregister\pageref7
\soulregister\section7
\soulregister\label7
\soulregister\begin7
\soulregister\centering7
\soulregister\caption7

\soulregister\tabcolsep7
\soulregister\begin7
\soulregister\textwidth7
\soulregister\hline7
\soulregister\end7

\usepackage{url}

\usepackage{algorithmic}
\usepackage{bibspacing}

\usepackage{graphicx}
\usepackage{textcomp}
\def\BibTeX{{\rm B\kern-.05em{\sc i\kern-.025em b}\kern-.08em
    T\kern-.1667em\lower.7ex\hbox{E}\kern-.125emX}}
\begin{document}
\history{Received November 1, 2018, accepted November 14, 2018, date of publication xxxx 00, 0000, date of current version December 31, 2018.}
\doi{10.1109/ACCESS.2018.2883151\\ \centering\textcolor{blue}{\copyright 2018 IEEE.  Personal use of this material is permitted.  Permission from IEEE must be obtained for all other uses, in any current or future media, including reprinting/republishing this material for advertising or promotional purposes, creating new collective works, for resale or redistribution to servers or lists, or reuse of any copyrighted component of this work in other works.}}

\title{Low Power Wide Area Networks: A Survey of Enabling Technologies, Applications and Interoperability Needs}
\author{\uppercase{Qahhar Muhammad Qadir}\authorrefmark{1}\authorrefmark{2}\IEEEmembership{Member, IEEE}, 
\uppercase{Tarik A. Rashid}\authorrefmark{1}\authorrefmark{3} \IEEEmembership{Member, IEEE}, 
\uppercase{Nawzad K. Al-Salihi}\authorrefmark{1} \IEEEmembership{Member, IEEE}, \uppercase{Birzo Ismael}\authorrefmark{1},
\uppercase{Alexander A. Kist}\authorrefmark{4} \IEEEmembership{Senior Member, IEEE} and {Zhongwei Zhang}\authorrefmark{5}
\IEEEmembership{Member, IEEE}}

\address[1]{Department of Computer Science and Engineering, University of Kurdistan Hewl\^{e}r, Erbil 44001, Iraq }

\address[2]{Department of Electrical Engineering, Salahaddin University-Erbil, Erbil 44001, Iraq}

\address[3]{Department of Software and Informatics Engineering, Salahaddin University-Erbil, Erbil 44001, Iraq}

\address[4]{School of Mechanical and Electrical Engineering, University of Southern Queensland, Toowoomba, QLD 4350, Australia }

\address[5]{School of Agricultural, Computational and Environmental Sciences, University of Southern Queensland, Toowoomba, QLD 4350, Australia }

\markboth
{Q. M. Qadir \headeretal: LPWA Networks: A Survey of Enabling Technologies, Applications, and Interoperability Needs}
{Q. M. Qadir \headeretal: LPWA Networks: A Survey of Enabling Technologies, Applications, and Interoperability Needs}

\corresp{Corresponding author: Qahhar Muhammad Qadir (e-mail: qahhar.qadir@ukh.edu.krd, qahhar.qadir@su.edu.krd)}

\begin{abstract}
Low power wide area (LPWA) technologies are strongly recommended as the underlying networks for Internet of things (IoT) applications. They offer attractive features, including wide-range coverage, long battery life and low data rates. This paper reviews the current trends in this technology, with an emphasis on the services it provides and the challenges it faces. The industrial paradigms for LPWA implementation are presented. Compared with other work in the field, this survey focuses on the need for integration among different LPWA technologies and recommends the appropriate LPWA solutions for a wide range of IoT application and service use-cases. Opportunities created by these technologies in the market are also analyzed. The latest research efforts to investigate and improve the operation of LPWA networks are also compared and classified to enable researchers to quickly get up to speed on the current status of this technology. Finally, challenges facing LPWA are identified and directions for future research are recommended. 

\end{abstract}


\titlepgskip=-15pt

\maketitle

\section{Introduction}
\label{sec:introduction}
\PARstart{I}{nternet of Things} (IoT) technologies have improved the way we live. These technologies address many of the challenges that humans are facing today, such as population growth, energy concerns and increasing demands for better means of sensing our environment. Approximately 28 billion devices, including more than 15 billion machine-to-machine (M2M) and consumer electronic devices, are expected to be communicating over short-range radio technologies such as Wi-Fi and Bluetooth and particularly over wide area networks (WANs) based on cellular technology by 2021 \cite{Ericsson2016} \cite{Sanchez-Iborra2016}. This is mainly due to the continuous decrease in the cost of sensors and actuators as well as innovations in communication technologies. We refer interested readers to \cite {Al-Fuqaha2015}, \cite{Islam2015}, \cite{Perera2014} and \cite{Atzori2010} for more details on the IoT technologies and applications.
	
Low power wide area (LPWA) networks have attracted considerable attention from the research community and industry as strong potential solutions for satisfying the requirements of diverse IoT applications. Although the term LPWA is relatively new, the design goals of this technology have been pursued for some time under different terms, including M2M, wireless sensor networks (WSNs) and IoT. IoT applications are characterized by their low data rates, power consumption and cost. Various sectors, such as transportation, healthcare, agriculture and industry are expected to exploit the features of LPWA technology. Personal healthcare systems, smart cities, smart grids, on-street lighting control and metering systems are among the applications that require communications over a large geographical area based on cheap and low-power devices. Such devices can be deployed and moved around over a wide area with the support of LPWA networks. Figure \ref{fig:lora} shows the main sectors in which LPWA technology can be deployed.

\Figure[t!](topskip=0pt, botskip=0pt, midskip=0pt){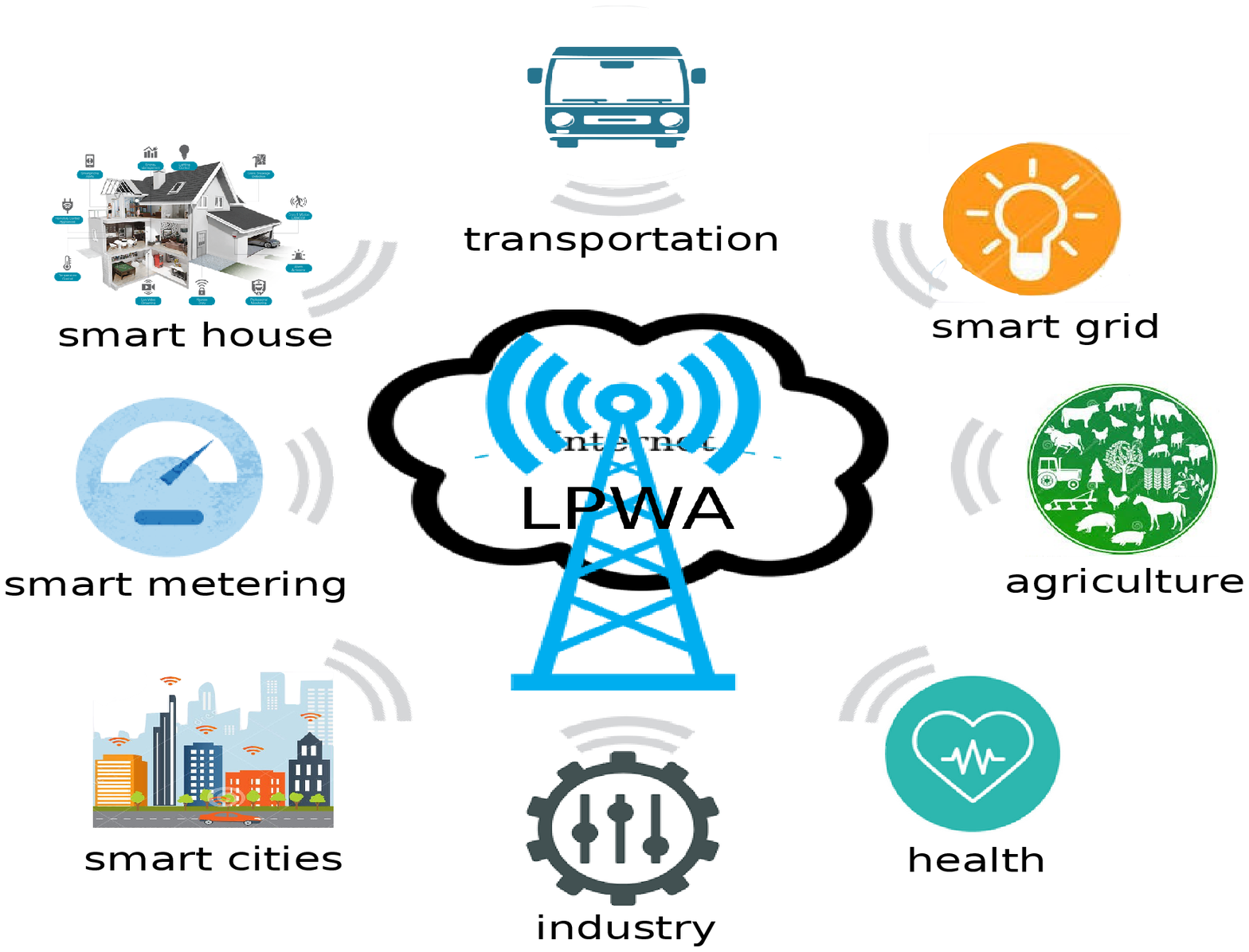} 
{Potential LPWA application use-cases \label{fig:lora}} 

\begin{figure}[t]
	\centering
	\includegraphics[width=\columnwidth]{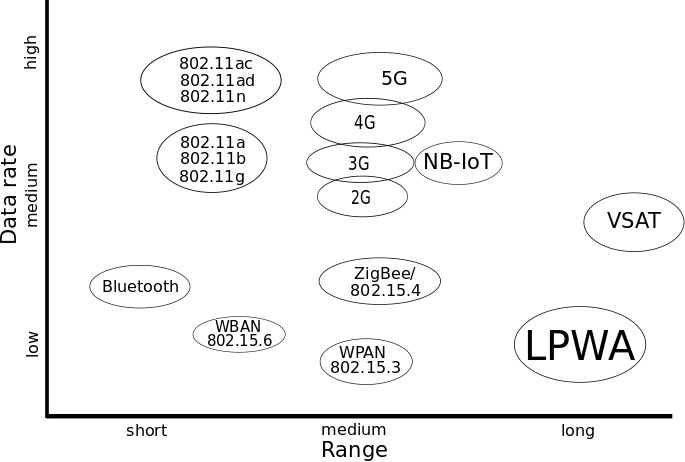}
	\caption{Ranges and data rates of wireless technologies \cite{Linklabs} \cite{Wang2017a}}
	\label{fig:lpwaTech}
\end{figure}

The main target of the LPWA technology is IoT applications that run on affordable low-battery devices and require communications over a wide geographical area. The success of such IoT applications, however, is associated with the limitations (in terms of data rate, number of devices and transmission range) of legacy wireless technologies. IoT applications traditionally operate over short-range wireless networks, e.g., ZigBee, Bluetooth, and Z-Wave; wireless local area networks (WLANs), e.g., Wi-Fi; and cellular networks, e.g., the global system for mobile communications (GSM) and long-term evolution (LTE). LPWA can be regarded as the outcome of efforts either to extend the range of WLANs and low-power wireless personal area networks (LoWPANs) or to minimize the cost and power consumption of cellular networks. Figure \ref{fig:lpwaTech} shows the ranges and data rates of various wireless technologies.
	
\begin{figure}[t]
	\centering
	\includegraphics[width=\columnwidth]{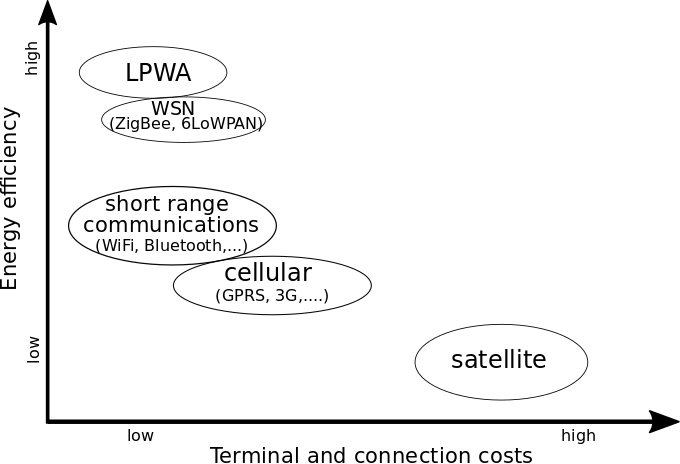}
	\caption{Energy efficiency and terminal and connection costs of wireless technologies \cite{Sanchez-Iborra2016}}
	\label{fig:energyCost}
\end{figure}

Non-cellular wireless technologies cover a few hundred meters at most, whereas traditional cellular technologies require devices to consume considerable energy \cite{Raza2017}. Modern cellular network architectures, such as 5G, offer improved energy efficiency for device-to-device (D2D) communication and multihoming, but challenges remain for low-power and low-cost devices that run IoT applications \cite{Singhal2016}. An LPWA network is defined as the low-power version of a cellular network, with each single cell covering thousands of end devices \cite{Augustin2016}. The energy efficiencies of various wireless technologies and their terminal and connection costs are illustrated in Figure \ref{fig:energyCost}. Technology sunsetting (the evolution of cellular generations) is another reason why cellular technologies are not sufficiently practical for IoT devices \cite{Ingenu}. 
	
Similar efforts have previously been made to review the LPWA literature in \cite{Raza2017}, \cite{Vangelista2015}, \cite{Sanchez-Iborra2016}, \cite{Sinha2017}, \cite{Wang2017a}, \cite{Mekki2018} and \cite{Song2017}. Both \cite{Sanchez-Iborra2016} and \cite{Raza2017} have explored the advantages and disadvantages of LPWA solutions as well as efforts to standardize the technologies; the current deployment status in Spain is also discussed in \cite{Sanchez-Iborra2016}. \cite{Raza2017}, on the other hand, is a more comprehensive survey. It mainly focuses on the goals of LPWA technology and the techniques used to achieve these goals. It also presents the efforts made by standard developing organizations (SDOs) such as IEEE, ETSI, 3GPP, and IETF as well as industrial consortia such as the Weightless SIG \cite{weightlessSIG}, the LoRa Alliance \cite{loraalliance}, and the DASH7 Alliance \cite{dash7} to standardize LPWA technologies. As three major low-power, long-range M2M-enabling technologies, LPWA, IEEE 802.11ah \cite{7920364}, and cellular-based M2M communication networks are compared in \cite{Wang2017a}. The work presented in \cite{Song2017} is limited to the review and comparison of LoRa and narrow band-IoT (NB-IoT) with general packet radio service (GPRS), while \cite{Sinha2017} and \cite{Mekki2018} are merely comprehensive surveys limited to SigFox, LoRa and NB-IoT, and the focus of \cite{Vangelista2015} is LoRa \cite{Sornin2017}.
	
Although this paper offers distinctive contributions, LPWA is still considered a new area of research, and therefore, the need for similar work is justified. It surveys the current trends in the LPWA networks. Through this survey, we present the most dominant LPWA technologies and the needs for integration among them and also address ways that LPWA technology can support the rapid growth of IoT applications and the challenges associated with these technologies. 

Compared with other surveys, our paper justifies the need for integration among different LPWA solutions and recommends suitable technologies for diverse IoT application and service use-cases. Short-range and long-range M2M wireless enabling technologies are compared and analyzed for the IoT. Moreover, recent research undertaken on all aspects of the LPWA is compared and categorized, and contributions and techniques are identified. The market opportunities inspired by these technologies are analyzed. This review provides researchers with a thorough survey to get them up to speed on the discussed topics in a single article. It is therefore an informative and up-to-date survey for readers interested in the LPWA networks. 
	
The remainder of this paper is organized as follows: Enabling technologies of short-range and long-range wireless M2M are compared and analyzed for the IoT in Section \ref{Sec:M2M}. Section \ref{Sec:marketOp} analyzes the market opportunities created by the LPWA industry. The need for integration among diverse LPWA technologies is justified in Section \ref{Sec:integrationNeed}. Section \ref{Sec:useCases} recommends suitable LPWA technologies for a wide range of IoT application use-cases. The main LPWA challenges are identified, and recommendations for future research are presented in Section \ref{Sec:LPWAchallenges}. The paper is concluded in Section \ref{Sec:Conclusion}.
	
\section{Wireless Technologies for M2M Communications}
\label{Sec:M2M}
Fast growing interest in M2M communications and unsuitability of wireless technologies designed for human use, such as cellular networks \cite{Xiong2015}, for this type of communication have pushed the industry to develop a number of wireless network solutions specifically for M2M communications. Devices through an M2M wireless system can collect information and share it with other devices to monitor or control the environment around us without the need for human intervention. 

In this section, we survey the dominant M2M wireless technologies and categorize them into short and long range solutions. We then analyze the trade-off in wireless systems for the IoT in terms of power and energy consumption, licensing, signal-to-interference-plus-noise ratio (SINR), data rates, reliability and end device active time. 

	\begin{table*}[!th]
		\centering
		\caption{Technical features of ZigBee, Bluetooth and WiFi }
		\label{tab:shortTechSum}
		\setlength{\tabcolsep}{5pt}
		\begin{tabular*}{\textwidth}
			{p{120pt}p{85pt}p{155pt}p{120pt}}
			
			\hline
			&	ZigBee	& Bluetooth	& WiFi generations\\
			\hline
			\hline
			Frequency	& 2.4 GHz \cite{Ahmed2016}	& 2.4 GHz \cite{BluetoothSpec}  & 2.4 GHz, 5 GHz \cite{Fehri2018}\\
			Channel access & CSMA/CA \cite{Ahmed2016} & FDMA/TDMA \cite{BluetoothSpec}& CSMA/CA \cite{Feng2017}\\
			Modulation &BPSK/OQPSK \cite{ieee802.15.4}&GFSK, $\pi/4$ DQPSK, 8DPSK \cite{Bluetooth}& Various \cite{Feng2017}\\
			Data rate (Maximum) &250 kbps \cite{Ahmed2016}&3 Mbps \cite{Bluetooth}& 7 Gbps \cite{Feng2017}\\
			Range (Maximum) & 100 m \cite{Ahmed2016} \cite{Ali2017}& 30m \cite{Ahmed2016}& 100 m \cite{Song2017}\\
			Connected devices (Maximum) &255 \cite{Song2017} &Over 1,000 in Bluetooth mesh networking \cite{Bluetooth} & 255 \cite{Song2017}\\
			
		\end{tabular*}
	\end{table*}

\subsection{Short-Range M2M Wireless Technologies}
\label{subsec:shortM2M}
There are a number of short-range M2M wireless technologies with different features and performances. Bluetooth, ZigBee, and WiFi are examples of these systems. The technical features and applications of these systems in the area of IoT are investigated in the sections below:

\subsubsection{ZigBee}

ZigBee \cite{ZigBee2008} is a specification designed for connecting low power wireless personal devices located in a small area. It features low data rates (a maximum of 250 kbps) and low power, which best fit applications that are delay tolerant. In addition to the low data rate, network range and capacity are two main limitations of ZigBee. ZigBee can connect up to 255 devices within a maximum of 100 m \cite{Song2006} \cite{Ali2017}.

ZigBee has been widely used in WSN for a wide range of applications including home and commercial building automation, industrial plant monitoring, fitness, wellness and intensively in health and aging population care \cite{Prabh2016}. It has been identified as a suitable solution in agriculture and environmental monitoring \cite{Ahmed2016}.

ZigBee uses the IEEE 802.15.4 \cite{ieee802.15.4} medium access control (MAC) protocol for MAC layer operation \cite{Prabh2016}. As a low-power consumption standard, IEEE 802.15.4 is not suitable for IoT applications that require coverage of large areas and communication among a large number of devices \cite{Ahmed2016}. Its small coverage range of 100 m and connectivity support for 255 devices can be achieved through operating in the unlicensed 2.4 GHz band using the direct sequence spread spectrum (DSSS) at data rates of as much as 250 kbps. 802.15.4 relies on the popular carrier-sense multiple access with collision avoidance (CSMA/CA) MAC protocol. 

\subsubsection{Bluetooth}
Portable devices can also be connected using a wireless technology standard called Bluetooth \cite{Bluetooth}. Bluetooth is an industry specification currently managed by the Bluetooth special interest group (SIG).
It sends small packets over multiple 1 MHz channels of bandwidth and uses short-range radio frequency (RF) from 2.402 to 2.480 GHz of the industrial, scientific and medical (ISM) band. 

The IEEE standard for Bluetooth is 802.15.1 \cite{BluetoothIEEE}, in which the radio for the physical layer and logical link control and adaptation protocol (L2CAP), link manager protocol (LMP) and baseband for the MAC layer are defined. Bluetooth has been mainly used for connecting wireless devices distributed in a small area (maximum of 30 m). Multiple radios allow a number of different applications such as streaming audio between a smartphone and speaker, controlling medical devices from a tablet or exchanging messages between nodes. On the one hand, the Bluetooth low energy (BLE) radio is designed for ultra low power and reliable and secure operations of continuous data streaming applications. The data are transmitted over 40 channels using a robust frequency-hopping spread spectrum (FHSS) to provide ranges for data rates (between 125 kbps and 2 Mbps), power levels (from 1 mW to 100 mW) and security options. The Bluetooth basic rate/enhanced data rate (BR/EDR) radio on the other hand, which is typical for burst data transmission, can support data rates from 1-3 Mbps and similar power and security levels of the LE radio \cite{Bluetooth}.

Bluetooth has been adopted by a number of use-cases including control and monitoring of smart building and industry and automation of heating, air conditioning, security, lighting and location services such as indoor navigation, asset and item tracking, space utilization and point of interest information \cite{Bluetooth}. BLE was identified as the most suitable short-range
communications solution in healthcare \cite{Baker2017}.

\subsubsection{WiFi}
IEEE 802.11 (also called WiFi) is another short-range technology that enables communication between wireless devices within a limited distance. WiFi is commonly used for closed environments such as homes and offices. It uses 5 GHz and 2.4 GHz in the ISM band. In addition to the IEEE 802.11, there are 802.11a/b/e/g/h/i/k/n/p/r/ac/ad/ax standards \cite{Feng2017}. Additional features were added to 802.11ac to improve the performance and speed and better manage the interference. It achieves these features through more channel bounding and MIMO and denser modulation.

WiFi has advantages over ZigBee and Bluetooth in terms of the data rate and coverage distance. New generations of WiFi allow nodes to communicate at very high data rates (e.g., 802.11ad has a data rate up to 7 Gbps) and, compared with ZigBee and Bluetooth, have much lower latency and higher power consumption. WiFi has managed to extend the communication distance and decrease delay (at the expense of power consumption), but the number of supported devices has remained a challenge \cite{Song2006} \cite{Khorov2015}. The WiFi improvements, therefore, do not meet the requirements of the IoT applications.

The technical features of the discussed short-range M2M wireless technologies are summarized in Table \ref{tab:shortTechSum}.

Although, ZigBee, Bluetooth and WiFi have been used for connecting various wireless devices, they are not able to address the demands of current IoT applications in terms of network range, capacity and power efficiency. This necessitates the design of other technologies that maintain the good features of available technologies and enhance their limitations. Furthermore, they should allow M2M communications while meeting the requirements of different IoT applications and can be deployed in all environments. 

\subsection{Long-Range M2M Wireless Technologies}
\label{Sec:LPWAtech}
The main requirements of the IoT applications are long distance coverage, high network capacity, low data rates, low power consumption, and affordable devices. The technologies discussed in Subsection \ref{subsec:shortM2M} are limited by the transmission range and number of devices. LPWA technology is considered a potential solution for M2M communications. M2M limitations on coverage and inefficient energy consumption force IoT providers to rely on LPWA networks for diverse IoT applications. 

The IEEE 802.11ah Task Group has been assigned by the IEEE 802 LAN/MAN standards committee (LMSC) to work on extending the range and data rate of the IEEE 802.11ah standard so that it achieves energy efficient protocols suitable for IoT applications \cite{Khorov2015}. IEEE 802.11ah is a WiFi based low data rate and low-energy consumption solution that can cover up to 1 km with 200 mW default transmission power at a minimum data rate of 100 kbps operating in frequency bands less than 1 GHz (TV White Space bands are excluded) \cite{7920364}. Using efficient modulation and coding schemes (MCSs) and proper propagation characteristics and operating in relatively narrow bands enable this technology to provide hundreds of Mbps (subject to channel conditions) and to be energy efficient \cite{Khorov2015}. The IEEE 802.11ah physical (PHY) layer is based on the IEEE 802.11ac and adopted to the use of sub bands. It uses orthogonal frequency division multiplexing (OFDM), multiple-input multiple-output (MIMO) downlink multi-user MIMO (DL MU-MIMO) and MCSs similar to IEEE 802.11ac. To cover long distances between a large number of low power nodes, the PHY and MAC layers implement a number of innovative approaches including hierarchical association identification, restricted access window, traffic indication map and segmentation, target wake time, and smaller headers \cite{Ahmed2016} \cite{Khorov2015}. 

IEEE 802.11ah is a promising communication solution for outdoor WiFi devices; however, it cannot be deployed in environments such as remote or underground areas \cite{Xiong2015}. LPWA networks have been proposed to replace short-range M2M wireless technologies. They have emerged to close the gap between local wireless and mobile wide area network technologies. They have features that are particularly attractive for IoT devices and applications, which will enable them to play a significant role over the next few years. LPWA technologies utilize various mechanisms to satisfy the requirements of modern IoT applications. Main characteristics of the LPWA networks are summarized below:

\begin{itemize}

	\item Long-range coverage
	
	LPWA networks offer optimized coverage for long-range communications, with a +20 dB gain over legacy cellular networks. This allows end devices to stay connected to their base stations over tens of kilometers \cite{Raza2017} (Figure \ref{fig:lpwaTech}).\\

	\item Low data rates
	
	LPWA networks target services that are not sensitive to losses and delays. For most of the use-cases presented in Section \ref{Sec:useCases}, single signals of a few bits are usually reported, which, in most cases, correspond to an "OK" status (Figure \ref{fig:lpwaTech}).\\

	\item Low power consumption
	
	Power consumption is a major concern with respect to mobile end devices. LPWA networks are designed to prolong battery lifetimes by placing devices in low-energy sleep mode and only waking them when they need to communicate with the gateway (Figure \ref{fig:energyCost}). This makes it possible for devices to operate on a single coin cell for several years \cite{Thubert2017}. LPWA applications operate over power-optimized radio networks.\\
	
	\item Low-cost end devices
	
	Since complex processing is mainly offloaded to the base station and the end devices are awake only when they have data to report, the cost of such a device can be as low as ten dollars or even less (Figure \ref{fig:energyCost}).	SigFox's packet would have been 15\% longer if  the message was decoded on the end devices \cite{Margelis2015}. For example, LoRa wide area network (LoRaWAN) and SigFox devices cost approximately $\$$2-5 each.\\

	\item Large numbers of end devices
	
	Two of the design goals for LPWA networks are high network capacity and scalability. Billions of devices are predicted for IoT deployment scenarios \cite{Ingenu}, in which a base station may be connected to hundreds of thousands of devices \cite{WeightlessSig2016}. LPWA technologies use a number of techniques to support such massive numbers of devices. Some of these techniques include the use of ultra narrow band (UNB) communication, which enables the coexistence of a large number of devices within a cell while controlling interference and adaptive data rates and channel selection \cite{Raza2017}. \\

	\item Usage of unlicensed spectrum
	
	Most LPWA technologies use free licensed spectrum resources within the ISM band. Therefore, LPWA providers do not need to pay for licensing, thus lowering the cost of deployment.\\

	\item Simplified network topology
	
	LPWA networks are deployed in a simple star topology (instead of the mesh or tree topologies used by other technologies), in which end devices are directly connected to the base station through their modems. This simple deployment scheme allows devices to operate over a thin and scalable infrastructure that spans a range of several kilometers. This also simplifies the installation of hardware. Additional base stations can be used to further extend the coverage. A star-of-stars topology can also be used to connect a cluster of base stations via gateways or concentrators \cite{Sanchez-Iborra2016}.
\end{itemize}

There are currently a number of LPWA providers, such as Semtech \cite{Semtech}, SigFox \cite{SigFox}, Ingenu \cite{Ingenu}, Silver Spring \cite{silverspring} and Telensa \cite{Telensa}. The diversity in the LPWA technologies is mainly due to the difference in the PHY and MAC specifications. The SDOs have exerted much effort to enable communication between these diverse technologies through standardizing or amending the design of the PHY and MAC layers. Some (e.g., Telensa) have already begun standardization of their LPWA technologies \cite{Raza2017}. The PHY/MAC challenges of SigFox, LoRa, Weightless and RPMA for the IoT have been analyzed in \cite{Goursaud2015}. In this section, we explore LoRaWAN, SigFox, RPMA \cite{rpma}, Telensa and NB-IoT \cite{3GPPTR36}. We refer readers interested in other LPWA technologies to \cite{Raza2017} and \cite{Sanchez-Iborra2016}.

	\subsubsection{LoRaWAN}
	\label{Sec:LoRaWAN}
	
	LoRa \cite{Semtech} stands for Long-Range, a physical-layer technology for long-range, low-power wireless communication systems invented by Semtech. LoRa, among other LPWA technologies such as SigFox and WiSUN \cite{wisun2012}, is one of the most promising and widely adopted technologies \cite{Adelantado2017}. Its robustness to interference and long-range coverage (more than 10 km) are made possible through the use of M-ary frequency-shift keying (FSK) modulation (symmetric for uplink (UL) and downlink (DL)) and chirp spread spectrum (CSS) modulation, a technique in which the signal is modulated by chirp pulses \cite{Liao2017}. Real experiments by \cite{Petajajarvi2015} have shown that LoRa can cover more than 15 km on the ground and close to 30 km on the water using the 868 MHz ISM band with a transmit power of 14 dBm and the maximum spreading factor (SF). LoRa is now available from mobile network providers around the world \cite{Semtech}.
	
\begin{figure}[t]
		\centering
		\includegraphics[width=\columnwidth]{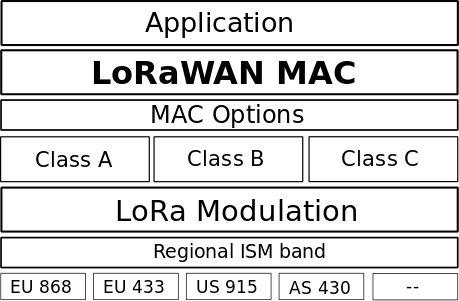}
		\caption{LoRa and LoRaWAN \cite{Semtech}}
		\label{fig:loraANDloraWAN}
	\end{figure}

	LoRa technology operates under the LoRaWAN protocol developed by the LoRa Alliance. LoRaWAN is designed for battery-based end devices. The end devices are connected to the central network server through a gateway (also called a concentrator or base station) in a star-of-stars topology. End devices make connections to one or more gateways using single-hop LoRa or FSK communication, whereas the gateways and network servers are connected using standard Internet protocol (IP) connections \cite{Sornin2017}. Figure \ref{fig:loraANDloraWAN} shows both LoRa and LoRaWAN.
	
	Different channels and data rates are utilized by end devices using unlicensed radio spectrum resources in the ISM band \cite{Sornin2017}. Studies have shown that sub-band selection and combination affect the quality of service (QoS) \cite{Soerensen2017}. An adaptive data rate scheme facilitates per end device data rates and frequencies and maximizes the network capacity and battery life of end devices. LoRa can support data rates from 0.3 kbps to 50 kbps \cite{Sornin2017}. To mitigate interference, end devices select channels in a pseudo-random fashion for every transmission. They also ensure that the maximum transmit duty cycle and maximum transmission duration are suitable to the sub-band used and compliant with local regulations \cite{Sornin2017}. It has been revealed that unlike channel duty cycle limitation, the duration of receive window affects the throughput for small packet size, SF has a significant impact on the network coverage and LoRaWAN is similar to ALOHA in that its performance degrades rapidly as load increases \cite{Augustin2016}. A near-optimal throughput was reported to be achieved by means of the pure ALOHA protocol and a proposed particle-filter-based retransmission control algorithm \cite{Seo2017}.

	Figure \ref{fig:loraANDloraWAN} illustrates how LoRaWAN supports various applications through three bidirectional classes of communication, as described below \cite{Sornin2017}:\\
	
	Class A allows end devices to communicate bidirectionally. This class is supported by all LoRaWAN devices and is compatible with Class B and Class C. Class A operation requires the lowest power end devices. The end devices receive random windows on the DLs. Pure ALOHA is deployed for ULs by devices in this class.
	
	Class B end devices have received windows at scheduled times in addition to the random receive windows of Class A. Beacon is used for time synchronization with the server and for determining the beginning of a time window. Applications that require additional DL traffic should run on Class B devices. Finally, Class C end devices have continuous open receive windows except during transmission. A LoRaWAN packet has an 8-symbol preamble, a header, a payload of 51-222 bytes and a cyclic redundancy check (CRC) code.
	
	\subsubsection{SigFox}
	\label{Sec:SigFox}
	SigFox \cite{SigFox} is another proprietary LPWA solution that operates in the 200 kHz band of the publicly available spectrum to transfer 100 Hz wide messages at a data rate of 100 bps or 600 bps. Using UNB modulation, SigFox achieves a range between 10 and 50 km and robustness against noise. The maximum payload sizes for UL and DL are 12 and 8 bytes, respectively. Thus, a message is expected to take only a short time (an average of 2 s) to reach the base station. Due to the small message size and a restriction on the number of packets per day, this solution requires less energy, consequently prolonging the battery life of devices. Each device is limited to a maximum of 140 UL messages and 4 DL messages under the strictest regulations \cite{Mekki2018} \cite{Baker2017}. SigFox relies on the cloud for computing and pushes the network complexity to the base station to achieve affordable and energy-efficient devices. Unlike in cellular protocols, a SigFox device can connect to any of the three base stations in its range. The resistance of SigFox to interference is based on the implementation of three different diversity schemes: time, frequency, and space \cite{SigFox}.
	
	\subsubsection{Random Phase Multiple Access (RPMA)}
	\label{Sec:RPMA}
	The proprietary LPWA technology known as RPMA \cite{rpma} was proposed by Ingenu (formerly On-Ramp). Unlike LoRa and SigFox, RPMA does not use sub-bands; it instead uses the global 2.4 GHz ISM band to realize a worldwide vision of the LPWA. RPMA relies on DSSS modulation, which does not impose a limit on the duty cycle or maximum frame duration in the global 2.4 GHz ISM band. For this reason, RPMA has the advantages of worldwide availability, wider spectrum utilization, and higher transmission power (and thus higher coverage) compared with LPWA technologies operating in fractured, regional sub-GHz ISM bands, such as LoRa and SigFox. RPMA features high data rates of up to 624 kbps on the UL and 156 kbps on the DL \cite{Wang2017a} \cite{Dhillon2017}, wider coverage and higher energy consumption than LoRa and SigFox due to the use of the 2.4 GHz ISM band \cite{Adelantado2017}. Ingenu designed RPMA \cite{Myers2013} for the physical layer to keep the packet size small and to enable unscheduled communication, which is not possible with DSSS \cite{Ingenu}. Interference in the 2.4 GHz ISM band and limitation on power in some regions such as Europe are the main drawbacks of this technology \cite{Goldhamme}.
	
	\subsubsection{Telensa}
	\label{Sec:Telensa}
	Telensa \cite{Telensa} uses UNB \cite{Massam2013}, a patent radio technology, to realize end-to-end communications for IoT applications. It operates in the license exempt radio spectrum, including ISM bands around the world. Telensa, in cooperation with ETSI, has made efforts to standardize their LPWA solutions. It features fully unicast bidirectional and broadcast communications that span over a wide range at low data rates. 
	
	\subsubsection{NB-IoT}
	\label{Sec:NB-IoT}
	A new LPWA solution called NB-IoT, which uses mobile network providers, was designed and standardized by 3GPP. Although NB-IoT is defined in Release 13 \cite{3GPPTR36} of the LTE standard, it is considered a standalone technology. NB-IoT was designed with IoT applications in mind, hence its name. NB-IoT relies on a decreased data rate (no higher than 158.5 kbps on the UL and 106 kbps on the DL \cite{Hoglund2017}) to reduce device cost and battery consumption. It therefore lacks LTE features such as measurements to monitor channel quality, carrier aggregation, and dual connectivity \cite{Sinha2017} \cite{Mekki2018}. NB-IoT is based on the core network of the LTE system, the evolved packet core (EPC) framework, defined in Release 8 of the standard. Under the EPC architecture, user data (also known as the user plane) and signaling (also known as the control plane) are separated to make the scaling independent \cite{3GPPEPC}. \cite{Miao2017} has found that NB-IoT has a lower UL time delay, higher channel utilization and wider coverage than the LTE technology. Handover in NB-IoT R13 is only possible prior to connection establishment \cite{Chen2017}.
	
	More features have been added to NB-IoT in 3GPP LTE Release 14 to provide better performance while maintaining the R13 current merits. Increased localization accuracy, higher data rates, offers for lower power consumption classes, improved non-anchor carrier operation, multicast and mobility support and coverage improvement are among these features \cite{Hoglund2017} \cite{Chen2017}.	The 3GPP Release 14-based commercial NB-IoT solution is now available \cite{NBIoTHuawei} \\

	Table \ref{tab:lpwaTechSum} summarizes the characteristics of the surveyed LPWA technologies.
	
	\begin{table*}[!th]
		\centering
		\caption{Characteristics of LoRaWAN, SigFox, RPMA, Telensa, NB-IoT \cite{Dhillon2017} \cite{Raza2017} \cite{Wang2017a} \cite{Sanchez-Iborra2016} \cite{SigFox} \cite{Semtech} \cite{3GPPTR36} \cite{RuanoLin2016} \cite{Sinha2017} \cite{Lin2016} \cite{Ingenu2} \cite{Hoglund2017} \cite{Baker2017} \cite{Mekki2018}   }
		\label{tab:lpwaTechSum}
		\setlength{\tabcolsep}{5pt}
		\begin{tabular*}{\textwidth}
			{p{45pt}p{85pt}p{105pt}p{75pt}p{75pt}p{69pt}}
			
			\hline
				&	LoRaWAN	& SigFox	&	RPMA	&	Telensa & NB-IoT\\
			\hline
			\hline
			
			Band	&Sub-GHz ISM: EU (433 MHz, 868 MHz), US (915 MHz), AS (430 MHz)	
			&Sub-GHz ISM: EU (868 MHz), US (902 MHz)  &2.4 GHz ISM	
			&Sub-GHz ISM: EU (868 MHz), US (915 MHz), AS (430 MHz)	 
			&Licensed 7-900 MHz\\		
			\hline
			
			Data rate	&LoRa: 0.3-5 kbps	&UL: 100/600 bps &UL: 624 kbps &UL: 62.5 bps	 &UL: 158.5 kbps \cite{Hoglund2017}\\
			
			&FSK: 50 kbps	&DL: 600 bps &DL: 156 kbps	&DL: 500 bps	 & DL: 106 kbps \\		
			\hline
			
			Occupied bandwidth	&125 kHz	&UL: 100/600 Hz, DL: 1.5 kHz &1 MHz	&100 kHz	 &200 kHz \\
			\hline
			
			Modulation	&LoRa/FSK	&UL: UNB/DBPSK, DL: GFSK	&RPMA/DSSS	&UNB 2-FSK	&QPSK \\
			\hline
			
			Range	&5 km (urban), 15 km (rural)	&10 km (urban), 50 km (rural) & 15 km (urban),	500 km	line of sight \cite{Dhillon2017} \cite{Wang2017a} &3 km (urban), 8 km (rural)	 &15 km\\		
			\hline
			
			Link budget (dB)	&151 (in EU), 171 (in US)	&146-162 &180 (FCC), 168 (in EU)	&Not known	 & 164  \\		
			\hline
			
			MAC	&Pure ALOHA	&R-FDMA &CDMA-like	&Not known	 &FDMA/OFDMA \\
			\hline
			
			Topology	&Star-of-stars	&Star &Star/Tree	&Star/Tree	 &Star \\		
			\hline
			
			Max. payload size (bytes)	&250	&UL: 12, DL: 8 &64	&65 k	 &13 \\
			\hline
			
			Link symmetry	&Yes	&No &No	&No	 &No \\
			\hline
			
			Handover	&Yes		&Yes	 &Yes	&Yes	 &Yes (in R14) \\
			\hline
			
			Error correction	&CRC-8/16	&UL: CRC-16, DL: CRC-8 &CRC	&Yes	 &CRC \\
			\hline
			
			Security	&AES-128 b	&Encryption done at upper level&16 B hash, AES 128 b	&Yes	 &L2 security \\		
			\hline
			
		\end{tabular*}
	\end{table*}

\subsection{IoT Wireless Technologies' Trade-off}
Traditionally, there has been an interest in wireless technologies that can support high data rates and that offer reliable and low latency communications. This trend has best addressed the requirements of human users who are normally greedy for optimum services. The response for this need was the emergence of wireless technologies represented by short-range systems and long-range cellular systems. 

High demand for communications between wireless devices has led to technologies that suit the applications running on these devices without needing human intervention. The outcome was a number of technologies that best fit these applications regardless of their differences in features and performances.

The first generation of M2M wireless systems managed to realize the IoT. However, there are drawbacks that limit the achievement of this goal. Some of these missing features of short-range M2M wireless systems are long communication distance, low-power consumption and high scalability. There are conflicting goals that make the adoption of these systems impossible. For example, ZigBee is able to range less than 100 m for a maximum of 255 nodes.
To cover more area and support more devices, one solution is to have gateways or relays connected in mesh topology, which results in a more complex environment, requiring more sophisticated protocols. Furthermore, such architecture requires complex MAC protocols to reduce packet loss and the number of retransmissions. Moreover, these low data rate systems consume more energy per transmission bit. To overcome the limitations of short-range M2M wireless systems, a new generation of technology has been proposed that minimizes these differences and trade-offs toward better supporting the requirements of IoT applications.

There are many trade-offs in short- and long-range M2M wireless systems. For example, LPWA is characterized by low power consumption but wide network range and simple deployment management for an enormous number of low cost devices \cite{Raza2017} \cite{Thubert2017} \cite{Vangelista2015}. Different techniques are used by LPWA to achieve these conflicting features. These techniques include the use of a reliable Sub-GHz band, efficient modulation, simple star topology, duty-cycling, simple MAC protocols, and simplified end-devices. Table \ref{tab:lpwaTechSum} explains the full list of these techniques.

There has been an understanding that licensed wireless technologies are better than unlicensed. This conception may need to be reconsidered with new technologies that operate in unlicensed spectrum and can provide close or similar performance \cite{Motorol2007}. As is the case with most technologies, wireless systems have provided cost-performance frameworks, i.e., better performance for higher cost. In the past 30 years, substantial improvements to unlicensed technologies have resulted in considerably higher speeds through better modulation, compression, interference mitigation and efficient spectrum techniques \cite{Motorol2007}. Although unlicensed technologies have advantages such as operation in any geographical area free of charge, unlicensed band frequencies (e.g., 5.4 GHz and 5.8 GHz in US) are not exclusively available for use.

Licensed systems do not require complex mechanisms to handle spectrum crowding and interference. Systems with such link performance operate at higher SINR and throughput compared with unlicensed systems. On the other hand, systems operate in license exempt frequency bands have a higher collision probability and thus lower SINR and throughput due to highly likely subjection to interference caused by other communications over the same frequency.

Before the emergence of LPWA, the industry showed vigilance towards the unlicensed part of the spectrum regardless of the substantial improvements introduced to these systems. The concerns were based on the fact that licensed systems are well prepared against interference and have higher reliability and performance \cite{Motorol2007}.

The technical differences between licensed and unlicensed technologies have been diminished by LPWA \cite{Masini2017}. They are divided over licensed and unlicensed spectrum technologies. LoRa, SigFox, RPMA and Telensa are unlicensed spectrum based, while Nb-IoT is licensed spectrum-based technology. There have been efforts by a group of mobile operators to standardize the LPWA technologies on licensed spectrum \cite{Linklabs}.

Transmission of unlicensed wireless systems does not necessarily mean that devices can stay active for a long time. SigFox, Telensa and LoRa, for example, use the unlicensed spectrum but are restricted to a 1\% duty cycle in EU 868 ISM sub-band restricting device activity and channel utilization. It was found that each end device can have a maximum transmission time of 36 s/h \cite{Adelantado2017}. 

As opposed to licensed systems, unlicensed wireless solutions are generally featured by high power consumption. For instance, RPMA operates in unlicensed global band 2.4 GHz ISM and consumes high power due to the use of high spectrum band \cite{Adelantado2017}. The data rates of LPWA technologies are relatively low, resulting in more energy incorporated in each transmitted bit. This denotes that, being low power, LPWA is not necessarily a low-energy system.

In terms of reliability, most unlicensed LPWA technologies (LoRaWAN, SigFox and Telensa) can provide reliable and robust communication with their low power transmission over the Sub-GHz band, which is not severely affected by attenuation and congestion \cite{Raza2017}. NB-IoT is a licensed technology that performs at 158.5 kbps; RPMA performs at 624 kbps and is unlicensed. Thus, with respect to LPWA, licensed systems do not necessarily have higher data rates than unlicensed systems.

\section{Market Opportunities}
\label{Sec:marketOp}
Industrial consortia have shown tremendous interest in LPWA technology since its emergence a few years ago. Many technology providers, including the LoRa Alliance, SigFox \cite{SigFox} and Ingenu \cite{IngenuPage}, have launched LPWA products. Telensa \cite{Telensa}, another LPWA provider, has deployed more than nine million devices across 30 countries. Smart lighting, parking and tracking are their main successful use cases \cite{Telensa}. 
	
It is forecast that there will be 25.1 billion IoT units (mainly consumer applications) in 2021, with an investment of \$3.9 trillion, representing growth at 32\% CAGR from 2016 \cite{Gartner2017}. NB-IoT is expected to have a massive share in this growth by connecting over 3 billion devices by 2023 \cite{ciscoNews}. IoT applications are expected to earn a revenue of 4.3 trillion dollars by 2024 \cite{Berthelsen2015}.
	
\begin{figure}[t]
	\centering
	\includegraphics[width=\columnwidth]{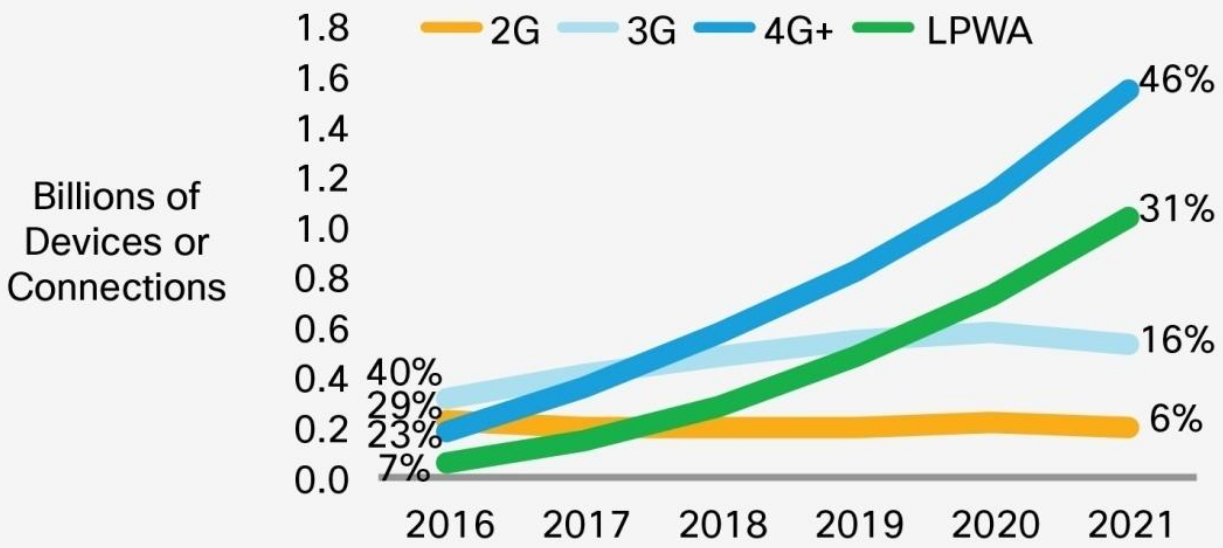}
	\caption{LPWA, 2G, 3G, and 4G+ (4G and 5G) global mobile M2M connections \cite{Cisco2017}}
	\label{fig:cisco2}
\end{figure}

Figure \ref{fig:cisco2} compares the percentage of the global share of LPWA M2M connections with the shares of 2G, 3G, and 4G+ (4G and 5G). It also shows the expected rapid growth in LPWA M2M connections, from 7$\%$ in 2016 to 31$\%$ by 2021 \cite{Cisco2017}. This trend indicates that mobile network operators are seeking other means of offering M2M connectivity to their customers as alternatives to cellular networks.
	
The expected growth of 4G technology depicted in Figure \ref{fig:cisco2} is due to its high data rates, low delays and strict security. These beneficial features, however, will not prevent mobile providers from deploying LPWA networks in the M2M segment. North America and Western Europe will be the top two regions for LPWA adoption by 2021, with shares of 31$\%$ and 20$\%$, respectively \cite{Cisco2017}.
	
\section{Need for horizontal integration among LPWA technologies}
\label{Sec:integrationNeed}
There are a number of LPWA technologies, each with common and unique features. The reason for this diversity in features and operations is that most LPWA solutions have been designed by the industry to fulfill certain market demands and have not gone through the SDOs. Many efforts are now being undertaken by IEEE, 3GPP, IETF and ETSI to bring these technologies closer together. A proper understanding of and compatibility among these diverse technologies will be necessary to make interoperability possible. Whereas gateways, backend base stations, IoT middleware and virtualization have been proposed to facilitate interoperability between LPWA technologies \cite{Raza2017}, we investigate the cause of this diversity and argue the need for interoperability between various technologies.

In this section, issues related to the compatibility among LPWA technologies are discussed, and the need to address this diversity and achieve better integration is explained. Diversity mainly arises due to the ways in which these technologies address the MAC and physical layers, more specifically, the bands they operate in, the modulation and MAC mechanisms, the payload formats and sizes, the communication modes, and the techniques used for forward error correction (FEC) and security.
	
LoRa and SigFox operate in unlicensed ISM radio bands, but each uses a different type of radio. SigFox's UNB uses a thin peak of the spectrum, whereas LoRa's CSS spans the entire available bandwidth \cite{Thubert2017}. By contrast, Telensa uses its own patented UNB radio technology operating in the license exempt spectrum. RPMA uses the global 2.4 GHz band, resulting in more interference and propagation loss at high frequencies. The robustness of physical layer enables this technology to operate in these challenging environments \cite{Vangelista2015}. Most LPWA solutions operate in the ISM band and thus have an advantage over NB-IoT, which operates in licensed cellular bands, because they do not require the involvement of mobile operators \cite{Adelantado2017}. This variety in radio technologies may be to the benefit of IoT applications in providing the service that fits best.
	
RPMA is not suitable for applications powered by battery due to the high processing power consumption of its underlying topology. SigFox, on the other hand, is not suitable for devices that require regular transmission due to its traffic limits, as mentioned in Section \ref{Sec:SigFox}. Moreover, SigFox's on-air transmission makes it non-compliant with some countries' regulations. The US, for example, imposes a maximum of 0.4 s of on-air transmission, which is much less than SigFox's approximate on-air transmission time of 2 s \cite{Sanchez-Iborra2016}. In addition, the band used in the US is less robust against interference than the band used in Europe. SigFox and Telensa lack the guarantee of QoS and thus limit the options for services demanding this feature. On the other hand, LoRaWAN and SigFox require a subscription to use cloud services \cite{Sinha2017} \cite{Finnegan2018}.
	
A possible issue that may arise from this diversity, particularly for solutions that are not standard-based, is complexity. Some of this complexity arises from the ways in which these technologies secure their communications and manage their services. An additional layer above the radio layer for the sake of convergence, similar to the IP layer of the Internet, has been proposed to mitigate this complexity. IPv6 and the constrained application protocol (CoAP) \cite{Shelby2014} are two potential protocols for providing connectivity among LPWA devices regardless of their underlying radio technologies \cite{Thubert2017}. This would allow Internet-based services to be provided to applications running on end devices.
	
Each LPWA solution offers a variety of characteristics that may or may not exist in other solutions. For example, WiSUN \cite{wiSUN}, a global industry alliance aimed at the standardization of wireless technologies for IoT applications already supports IPv6 through the IETF IPv6 stack for IPv6-based LoWPAN (6LoWPAN), but additional work will be required on security and identity management, as IoT devices are protected differently from computers. The IoT-based attack on KrebsOnSecurity.com could serve as evidence of this threat \cite{Thubert2017}.
	
The PHY and MAC layers of RPMA are based on the IEEE 802.15.4k standard \cite{6581828}. The standard targets long-range and low-energy critical infrastructure monitoring networks. The IEEE 802.15.4k standard has been amended to adopt DSSS modulation in the PHY and MAC layers \cite{Wunsch2017}. It also provides a QoS guarantee by prioritizing traffic \cite{Raza2017}. Both SigFox and Telensa use UNB modulation, and LoRa uses the orthogonal sequence spread spectrum (OSSS) modulation technique. The goal of ETSI's low throughput network (LTN) technology is to standardize bidirectional low-rate LPWA technologies. According to the LTN specifications \cite{ltnSpecifications}, either the UNB or OSSS modulation scheme can be deployed by LPWA solutions in sub-GHz ISM band. However, it is recommended to use binary phase shift keying (BPSK) on the UL and Gaussian frequency shift keying (GFSK) on the DL for UNB implementation and OSSS for bidirectional communications. The LTN approach has motivated a few LPWA providers, including SigFox, Telensa, and Semtech, to standardize their solutions while maintaining proprietary modulation schemes \cite{Raza2017}.
	
The data rate of LoRaWAN depends on the SF and bandwidth. To avoid collision, simultaneous transmissions must select different SFs and channels. In addition to PHY/MAC constraints, the performance of LoRaWAN is limited by duty cycle regulations in the ISM bands. A key constraint for LPWA technologies operating in unlicensed bands is the restriction imposed on the duty cycle, which limits an end device's occupation of the channel. SigFox, Telensa and LoRa devices are restricted to a maximum $1\%$ duty cycle in the EU 868 ISM band \cite{Adelantado2017}. The maximum transmission time for each end device operating in each EU 868 ISM sub-band was computed to be 36 s/h. Another issue that arises from the duty cycle restriction is that the channel selection must be compliant with the maximum duty cycle \cite{Adelantado2017}.
	
\section{LPWA Suitability for IoT application and service Use-Cases}
\label{Sec:useCases}

The characteristics discussed in Section \ref{Sec:LPWAtech} make LPWA networks suitable for IoT applications. A wide range of IoT services mainly require low data rate reporting, constrained by low-power end devices. A few times a day, small delay-/loss-insensitive frames, on the order of tens of bytes, are sent by these devices \cite{Thubert2017}. Examples of such IoT applications include measurement (such as of soil features), monitoring (such as of natural materials), and prediction of failure (such as of car engines). In most industries to date (except for oil and gas), the cost of deploying wired or cellular networks to collect and report data has been too high. LPWA networks enable the monitoring and reporting of this information in a cost-effective way due to the low cost of devices, deployment and operation. LPWA technology can support IoT applications with lower costs for a given number of monitored objects and a given range of deployment. 

LPWA networks are expected to be increasingly used in various sectors. LPWA feasibility has been demonstrated for infrastructure monitoring, transportation, asset tracking, security and health care \cite{Xiong2015}. LPWA has also been used for receiving data from a lighting monitoring system \cite{Myers2013}. The LPWA solutions that are investigated in this paper are recommended for a number of use-cases. Although LPWA technologies can be deployed in a wide range of sectors, we focus on the use-cases explained in Figure \ref{fig:lora}. Our recommendations are based on the features of the LPWA technologies that best meet the requirements of each use-case. We further limit the recommendations to the technical parameters listed in Table \ref{tab:lpwaTechSum}.
Therefore, the implementation details such as the architecture, hardware and software are beyond the scope of this paper. Interested readers are referred to \cite{Al-Fuqaha2015} \cite{Xiong2015} \cite{Baker2017} and \cite{Cao2016} for details on the IoT system architecture and implementation. 

\begin{table*}[!th]
	\centering
	
	\caption{Suitable LPWA technologies for major use-cases}
	\label{tab:useCases}
	\begin{tabular*}{\textwidth}
		{p{50pt}p{45pt}p{40pt}p{40pt}p{280pt}}	 		
		\hline
		
		Use-case	&	Bitrate & Mobility & Real-time & Suitable technology\\
		
		\hline
		\hline			
		
		Smart house & Low-High & No & Yes/No& Any for periodic reports and event-driven alerts, RPMA and NB-IoT for video surveillance monitoring\\
		Transportation & High & Yes & Yes & RPMA and NB-IoT\\
		Smart metering & Low-Medium & No & No& Any\\
		Smart grid & Medium & No & Yes& LoRaWAN, RPMA, NB-IoT\\
		Agriculture & Low-Medium & Yes & Yes/No & LoRaWAN, SigFox, Telensa for periodic reports and event-driven alerts, LoRaWAN for two-way real-time control\\
		Health & Low-High & Yes & Yes & LoRaWAN, RPMA, Telensa, NB-IoT for medical periodic report and event-driven alerts, LoRaWAN, RPMA, NB-IoT for two-way real-time medical control, RPMA and NB-IoT for real-time video based health conditions	monitoring\\
		Industry & Medium & No & Yes & LoRaWAN, RPMA and NB-IoT\\
		Smart cities & Low-Medium & Yes & Yes& SigFox, LoRaWAN, Telensa, NB-IoT for smart lighting and smart waste control and LoRaWAN, Telensa for smart parking and safe driving\\
		
		\hline

	\end{tabular*}
\end{table*}

To make the proper suggestion for each use-case, we must first understand what is required by each application. We characterize the use-cases of Figure \ref{fig:lora} according to their demands for bitrate, mobility and real-time communications. We take into account that other requirements such as low deployment costs, wide network coverage, high capacity, strong security and high energy efficiency are the design-to-criteria of LPWA, which are for the most part provided by the technologies listed in Table \ref{tab:lpwaTechSum}.

The suitability of LPWA technologies for real-time support is recommended based on a data rate of 28.8 kbps for real-time services, such as two-way control applications \cite{ITU-T2003}, and between 130 kbps to 4 Mbps with MJPEG coding (low quality) and  MPEG-4/H.264 coding (1920 x 1080 resolution and 30 fps) for IP-based video surveillance systems \cite{Adelantado2017}. Generally, LoRaWAN, RPMA and NB-IoT meet the data rate requirements of two-way control applications, while only the latter two satisfy the minimum data rate requirements of IP-based video surveillance systems. Although each LPWA technology can support regular reporting by sending short periodic messages, not all are suitable for frequently changing environments (ex: smart healthcare where reporting event-driven alerts is inevitable) such as SigFox due to local regulations and other operating restrictions. LoRaWAN with duty cycle regulation does not satisfy the requirements for deterministic monitoring and real-time operation \cite{Adelantado2017}. Unlike other technologies, NB-IoT operates within a licensed network, and the service is not available in some areas such as outlying districts and agricultural lands. LoRaWAN must be carefully dimensioned for each use-case based on key limiting factors such as the number of devices, the selected SF, and the number of channels \cite{Adelantado2017}. The comparative study in \cite{Mekki2018} found that SigFox and LoRa are efficient in terms of battery lifetime, network capacity and deployment cost, while NB-IoT has the advantage of low latency and support for QoS.

Smart meters with sensors monitor and meter the usage of resources such as oil, gas, electricity and water or other relevant information such as soil and oil pipeline status and report this information to the monitoring center. This information is sent periodically or when events that require urgent attention occur such as device breakdown. LPWA networks have yet to be widely used for operating sophisticated meters. Smart meters are no longer only used for measuring the amount of resources passing through them. They should be capable of tracking peak demand, measure utility quality and meter temperature, enable communication between multiple meters \cite{USDoEnergy2015} and even determine whether we are home \cite{Jin2017}. These meters are normally stationed on premises and require low reliability and low to medium bitrates \cite{Xiong2015} \cite{Masini2017}. Given these requirements, any studied technologies can be deployed taking into account the SigFox's daily limited number of packets and high latency, which may disqualify it for critical event-driven alerts. NB-IoT and RPMA both have preferences over other technologies for their long-range coverage and concerns and at the same time for their licensing requirements and power consumption. Wireless M-Bus is a European standard specified in EN 13757-5:2015 \cite{CEN2015} that is widely deployed in Europe for metering applications such as water, gas and electricity. LoRa and NB-IoT have been proposed for smart gas metering \cite{Dong2017} \cite{Chen2017}, water metering \cite{Chen2017} and electric metering \cite{Mekki2018} \cite{Adelantado2017} \cite{Hoglund2017}, whereas RPMA, SigFox, LoRaWAN and NB-IoT were recommended by \cite{Wang2017a}. LoRaWAN, SigFox, and RPMA were selected as suitable LPWA networks for smart metering applications \cite{Margelis2015}. Tens of thousands of electricity meters were set up with LoRa technology in the region of the Calenberg Land near Hanover, Germany \cite{Semtech2017}.

Smart house and industrial use-cases require automations such as monitoring and control at bitrate ranges from low to high for smart house applications and low for industrial applications. These applications are commonly stationed and thus do not require mobility support. Smart house applications, however, may require video surveillance that can be supported by RPMA and NB-IoT considering the video surveillance minimum data rate requirement (130 kbps) as mentioned earlier in this section. LoRaWAN was found to be inappropriate for video surveillance due to its low data rate \cite{Adelantado2017}. However, any LPWA technology can be deployed for periodic reports. NB-IoT was recommended by \cite{Chen2017} \cite{Hoglund2017} \cite{Song2017} and LoRa and SigFox by \cite{Mekki2018} for a similar purpose. 

On the other hand, industrial application requirements for two-way control automation can be met by LoRaWAN, RPMA and NB-IoT. It is worth mentioning that LoRa, SigFox and NB-IoT have been recommended by \cite{Mekki2018} for manufacturing automation. However, \cite{Adelantado2017} argues that LoRaWAN can only be a candidate solution for small networks with careful configuration of the SF and number of channels. KNX \cite{knx} and EN 50090-1:2011 \cite{EC2011} are two standards for home automation. Modbus series are a number of protocols used for automation and monitoring in industry \cite{Urbina2017}.  

Health care systems should collect real-time measurements \cite{Baker2017} of patients through relevant sensors, send them to health centers and prioritize critical conditions that require urgent attention. Furthermore, mobility support for remote clinics and patients is required. A number of licensed and unlicensed options are available for this use-case. Given that two-way real-time medical control requires at least 28.8 kbps \cite{ITU-T2003}, LoRaWAN, RPMA and NB-IoT can be used. However, remote doctors occasionally need to have video sessions with their patients. Thus, similar to smart house applications, RPMA and NB-IoT can support real-time video based monitoring of health conditions. Finally, any investigated technology is suitable for medical periodic reports and event-driven alerts except SigFox for the same reason mentioned above. Systems that are deployed for smart healthcare obviously should be energy efficient and support a large number of devices over a wide range. The best options are NB-IoT (overlooking the licensing costs) and LoRaWAN in terms of network capacity and range, with 53,547+ versus 40,000 nodes and a 15 km (urban) versus 5 km (urban) range, respectively. NB-IoT was determine to be the most suitable long-range communications solution in healthcare \cite{Baker2017}. The combination of short-range and long-range M2M communications is proposed for healthcare in which the central node of the wireless body area network (WBAN) and sensors are connected using technologies such as Bluetooth and ZigBee, whilst WBAN is connected to the provider's base station using a suitable LPWA technology \cite{Baker2017}. Although such a system is good in the sense that available short-range M2M architectures may be employed, low power consumption, scalability and less complexity are unlikely to be achieved.

LPWA supports devices equipped on vehicles or along the road in the transportation use-case to send road condition information such as car accidents and road congestion to the management center. They can also exchange information among themselves to organize traffic and increase road efficiency. LPWA technologies deployed in this environment must provide high throughput, reliable, real-time transmission of critical information, and real-time two-way control of mobile vehicles \cite{Masini2017} \cite{Xiong2015}. LoRaWAN is not a fit for this environment due to its delay and jitter constraints and contention caused by the ALOHA protocol. RPMA and NB-IoT can address the needs of smart transportation systems bearing in mind the high energy consumption of RPMA and the licensing cost of NB-IoT, for which trade-offs should be made. NB-IoT was found as an appropriate solution for vehicle and asset trading \cite{Hoglund2017}. 

LPWA technologies can also be utilized to monitor and control environmental and agricultural processes such as land watering and leak detection using small periodic or event-driven messages. Wide coverage areas and delay tolerance (except for reporting a fault or leak) are mainly the priorities of these applications. LoRaWAN, SigFox and Telensa can support delay tolerant messages for periodic reports and event-driven alerts. LoRa and SigFox have been recommended by \cite{Mekki2018}. However, SigFox and Telensa are not suitable for real-time control due to their limitations on data rates (less than 28.8 kbps, the minimum required data rate for these types of applications). NB-IoT is also not suitable for smart agriculture services due to poor availability of the cellular network in these outskirt areas. High power consumption of RPMA is a major challenge for remote nodes and therefore excludes this technology as well. These recommendations consider mobility a requirement of these applications \cite{Xiong2015}. LoRaWAN has been recommended for agriculture applications by \cite{Song2017} and \cite{Adelantado2017} with adequate deployment of gateways.

Smart cities are perhaps the most vital area of IoT, with an increasing number of applications and services. They bring a number of benefits to drivers, authorities and society. These benefits include reducing traffic by directing drivers to less congested roads and parking spots, reducing driving risks through restricting cars to traffic light control, reducing energy consumption by controlling street lights based on vehicle movement and managing waste intelligently. These applications are common in the sense that a huge daily number of small delay sensitive messages is triggered by events such as parking status in the case of smart parking, traffic jams in the case of smart traffic and garbage load in the case of smart waste management. In smart lighting, however, messages are sent simultaneously during sunrise and sunset, increasing the possibility of collision, while smart waste management systems can tolerate some latency. The variation of these needs impose deployments of more than one technology in this use, for example, a technology for smart waste control and another one for smart parking. However, LPWA technologies deployed for smart cities are required to cover a wide range and provide high security for a huge number of energy efficient nodes. SigFox, LoRaWAN and Telensa (in order) are appropriate license-free technologies for smart lighting and smart waste control as they meet the requirements mentioned earlier at the cost of some delay. The daily generated number of event-driven messages and high energy consumption disqualify both SigFox and RPMA and favor LoRaWAN and Telensa as two adequate solutions for event-driven alerts and two-way real-time control of smart parking and safe driving applications. LoRaWAN has been proposed for smart cities \cite{Adelantado2017} and successfully tested in real parking scenarios \cite{Song2017}. NB-IoT is also suitable for a number of smart city services \cite{Hoglund2017} at the cost of a licensing charge and is being deployed for smart city applications and services \cite{m1}. Although NB-IoT was selected for streetlights, parking and waste management by \cite{Chen2017}, the cellular signal is normally weak in underground car parking lots and remote areas, which shows that the NB-IoT is unsuitable for smart driving and parking. The IEEE 802.15.4k-based LPWA network has been proposed for air quality monitoring \cite{Zheng2016} and critical infrastructure monitoring \cite{Xu2017} and Telensa for smart cities \cite{Sanchez-Iborra2016}. Although LPWA-based air monitoring systems have advantages over traditional WSN-based systems such as wide monitoring coverage, data accuracy and real-time reporting remain challenging. This is due to the restrictions some of the LPWA technologies place on the number of packets each end node is allowed to send.

Smart grid is an IoT version of the legacy electricity network. It relies on bidirectional communications between the electricity supplier's distribution substation and consumer's electric meters to deliver and control the power in an efficient, optimized, safe, reliable and cost effective way. With smart grid, consumers' smart appliances are enabled to switch to other renewable energy sources such as solar during peak hours. In addition to the delivery of electricity, information is exchanged between the consumer's smart meter and supplier's controller to allow real-time control of power delivery \cite{Yan2013}. Monitoring that allows detection of malfunction or failure and control that warrants proper actions are two main requirements of smart grid applications. A large number of messages \cite{Aggarwal2010} are expected to be communicated between a large number of consumers and the supplier in addition to securing the vulnerable stationed system elements. The LPWA technology should also be scalable to allow joining of a growing number of new devices. Given these requirements, LoRaWAN, RPMA (considering power consumption is not a concern of smart grid stationed devices) and NB-IoT (overlooking the licensing cost) are convenient for smart grid applications.

Transportation, agriculture and remote healthcare applications are characterized by mobility of nodes, strong security measures, large network range and dense capacity. SigFox and Telensa can manage interference when the number of nodes grows using the UNB modulation as in LoRaWAN by adapting the data rates. On the other hand, smart metering requires a wide coverage area but not necessarily mobility. 

Based on our arguments, we recommend LPWA technologies to each of the discussed use-cases as listed in Table \ref{tab:useCases}.

\section{Challenges and future research directions}
\label{Sec:LPWAchallenges}

There have been extensive efforts by the research community to optimize or evaluate LPWA operation. Table \ref{tab:lpwaResearch} summarizes these contributions. These works, however, have not addressed all issues associated with the LPWA network systems. In this section, we discuss current LPWA challenges and present directions for future research in this field.

\begin{table*}[!t]
	\centering
	\caption{Summary of LPWA-related research activities}
	\label{tab:lpwaResearch}
	\begin{tabular*}{\textwidth}
		{p{10pt}p{85pt}p{95pt}p{50pt}p{210pt}}
		
		\hline
		
		Ref.	&	Contribution & Methodology & LPWA technology& Outcome\\		
		\hline
		\hline			
		\cite{Xu2017}	& Design/prototyping & Design, testbed & Not specified & System spans up to 3 km with concurrent end devices\\		
		\hline
		\cite{Augustin2016} & Performance evaluation, analysis & Field tests, simulations, testbed & LoRa & -LoRa has good resistance to interference\\ 
		&&&& -LoRa covers up to 3 km in suburban areas\\
		&&&& -LoRaWAN's performance similar to that of ALOHA, degrades quickly as load increases\\
		\hline
		\cite{Miao2017} & Review, performance evaluation & Modeling, simulation & NB-IoT & -NB-IoT has lower UL delay than LTE (less than 10 s)\\
		
		&&&& -Higher channel utilization than LTE\\
		&&&& -Wider coverage range than LTE\\
		\hline
		\cite{Soerensen2017} & Optimization of resource allocation, performance investigation & Analytical model, simulation & LoRaWAN & Sub-band selection and combination affect the QoS \\

		\hline			
		\cite{Adelantado2017} & Overview, performance evaluation, analysis & Application use cases& LoRaWAN& -LoRaWAN requires careful dimensioning for each use case\\
		&&&& -Example: current LoRaWAN does not guarantee monitoring and real-time applications\\
		\hline

		\cite{Georgiou2017} & Scalability, performance study &Modeling using a stochastic geometry framework& LoRa &-Exponential drop in coverage probability with an increasing number of end devices\\
		&&&& -Use of the same spreading sequence has a greater impact on scalability than spectrum restrictions\\
		&&&& -Co-spreading factor interference provides rigorous scalability\\
		\hline
		\cite{Bor2016} \cite{lorasim2016}& Scalability investigation & Experiment-based modeling& LoRa& LoRa scales well with a dynamic parameter configuration and/or multiple sinks\\
		\hline
		\cite{Petajajarvi2015} & Coverage measurements, empirical channel attenuation model development & Real-life experiments& LoRa& Up to 15 km ground coverage and 30 km water coverage measured for the 868 MHz ISM band, 14 dBm transmission power, and the maximum SF\\
		
		\hline
		\cite{Seo2017}& Retransmission control algorithm & Modeling and analysis&Not specified&Near-optimal throughput achieved\\  
		\hline
		\cite{Voigt2017}& Interference mitigation& Simulation&LoRa&Multiple base stations outperform directional antennas\\
		\hline
		\cite{Lieske2016}& Decoding performance study &Modeling and simulation& Not specified&FEC encoded time-hopping spread spectrum systems outperform single-packet systems\\
		
		\hline
		\cite{KIM2017}& Congestion and rate control scheme& Analysis&Not specified&Data efficiency of the proposed scheme\\
		\hline
		\cite{BARRACHINAMUNOZ2017153}& Energy-efficient optimal routing connection& Distance-ring exponential stations generator (DRESG) framework& Not specified&Bottleneck reduction and longer network lifetimes\\
		\hline
		\cite{Roth2017}&Analysis of Turbo-FSK, parameter optimization, influence of packet length on energy& EXIT chart&IEEE 802.15.4k, LoRa, TC& 
		
		Turbo-FSK achieves promising performance, less complexity and a constant envelope on the transmitter side \\
		\hline
		\cite{Liao2017}&Receiver performance evaluation and improvement, introduction of timing offsets between relaying packets &Field experiments&LoRa&LoRa combined with concurrent transmission provides a high packet reception rate\\
		
		\hline			
		
	\end{tabular*}
\end{table*}

In 2016, nearly 17.6 billion devices, including smartphones, tablets, and computers \cite{Nordrum2016}, were connected via IoT technologies, and this number is set to hit 25.1 billion by 2021 \cite{Gartner2017}. However, what if each of these 25.1 billion devices could be easily attacked and compromised by hackers? According to a Business Insider survey, 39$\%$ of respondents regard this possibility as a great concern. This concern is understandable because these devices collect and transmit data that are valuable and often sensitive and should be secured as such. Since LPWA communications mostly operate in the free unlicensed spectrum, LPWA-based IoT applications will then have security concerns. Although SigFox applies different techniques to secure the network, such as sequencing, message scrambling and anti-replay, it does not encrypt the transmitted payload, and thus it relies on the upper layer for this purpose. SigFox devices rely on a fixed secret key for online registration, which can be exploited for sending forge messages to the network. This may lead to a node being blacklisted if more than the limited number of messages is exceeded \cite{Margelis2015}. RPMA is also reported to be vulnerable to security \cite{Margelis2015}. LoRaWAN end devices, on the other hand, can be exploited when activated Over-The-Air through the nonce that is used to send an unencrypted join request \cite{Margelis2015}. These issues can be addressed by applying better security techniques or by operating in the licensed spectrum. Applying up-to-date security patches and protocols that allow device mobility while enforcing security such as the locator identity separation protocol (LISP) \cite{rfc6830} and network mobility (NEMO) \cite{rfc3963} are some of these techniques \cite{Thubert2017}.

The extensible authentication protocol method for 3rd generation authentication and key agreement (EAP-AKA) \cite{rfc4187} and EAP method for GSM subscriber identity modules (EAP-SIM) \cite{rfc4186} are designed for cellular devices while lightweight authentication protocols such as the pre-shared key EAP method (EAP-PSK) \cite{rfc4764} are proposed for devices that require little computing power and memory. A lightweight bootstrapping service is proposed for IoT devices in \cite{Garcia-Carrillo2016} using CoAP, EAP and authentication, authorization and accounting (AAA) architecture. Currently, works are in progress within IETF which encourage the use of EAP such as EAP-AKA$'$ \cite{EAP-AKA} (a small revision of EAP-AKA), EAP-transport layer security (EAP-TLS1.3) \cite{eap-tls13-02} and nimble out-of-band authentication for EAP (EAP-NOOB) \cite{EAP-NOOB}. EAP-TLS is an EAP authentication method which is widely supported in WiFi, MulteFire \cite{multefireSpecifications} and 5G networks. EAP-NOOB is proposed for the registration, authentication and key derivation of IoT devices that have a minimal user interface and no pre-configured authentication credentials, and could also be a candidate for authenticating LPWA devices due to similar device characteristics. On the other hand, IETF recommends AAA framework \cite{rfc2904} to tackle some of the security issues of LPWA networks \cite{rfc8376}, and it has been considered as one of the technologies to secure IoT deployments in \cite{Thubert2017}. Authentication of the massive number of nodes that number in the hundreds of thousands in some scenarios is a challenge for LPWA technologies. This is also true for data encryption \cite{Raza2017} \cite{Garcia-Carrillo2018} \cite{Garcia-Carrillo2017} \cite{Islam2015}. Furthermore, security mechanisms of systems that rely on the cloud for data storage and processing demand further investigation. This should not be overlooked for healthcare systems, where privacy is mandatory \cite{Al-Fuqaha2015} \cite{Islam2015} \cite{Baker2017}.

Some of the LPWA solutions are currently based on the cloud, and the future trend is likely towards more cloud dependence. The large amounts of data collected from sensors deployed in end devices must be processed, and information should be generated in a way that is meaningful to humans. A framework that provides mechanisms for data collection, analysis and resource provision in smart cities is presented in \cite{Santos2018}. Traditional approaches to data processing are no longer appropriate with respect to the IoT environment \cite{Baker2017}. This task requires extensive research that can lead to developing efficient and secure data mining algorithms and implementing machine learning for managing big data.
	
The LPWA technologies surveyed in this paper provide basic support for roaming, which allows network deployment on a large scale. This matter is being investigated by the LoRa-Alliance in the LoRaWAN 1.1 specification \cite{Adelantado2017}. SigFox makes it possible for devices to communicate with multiple base stations without the need for roaming or handovers \cite{Mekki2018}. However, mobility support must be further developed to include services such as end-to-end secure communication, inter-operator billing, device location and transparent provisioning throughout roaming periods.

Although LPWA uses a number of robust and reliable techniques, the massive number of LPWA devices that are mainly operating in shared radio spectrum are error-prone. This is a serious issue for data transmission (as is the case in the LPWA networks where short messages are often transmitted), which requires a high level of accuracy. Message integrity concerns the LoRaWAN DL as frames are not checked for CRC and NB-IoT supports a single adaptive asynchronous hybrid automatic repeat request (HARQ) for both links \cite{Hoglund2017}. This raises the issue of message integrity, which needs to be properly addressed.

The low-end device cost is a potential economic factor driving the deployment of LPWA systems by businesses and governments. This one-time cost, however, is not as significant as the cost of service subscription. Thus, LPWA operators must keep this cost as minimal as possible. 

On the one hand, the use of licensed spectrum resources by some LPWA technologies will increase their operational costs due to the license subscription requirement. On the other hand, the large number of objects, whose number will increase even more in the future, operating in the free unlicensed spectrum increases the probability of packet collision. \cite{Liao2017} has found that although LoRa is robust to packet collisions resulting from concurrent transmissions, introducing timing offsets between relaying packets to add a random timing delay can further improve the receiver performance. CSMA/CA or cellular like multiple access protocols seem to be very complicated for simple and cheap LPWA devices. Although the use of Sub-GHz band mitigates the possibility of interference at low power budgets and duty cycle regulations limits device activity on the network, more efficient protocols are vital for better managing access to the network \cite{Raza2017}. For LPWA technologies that operate in the licensed spectrum, the issues of frequency licensing and spectrum management must be addressed. There is a potential for new radio schemes to be introduced as needs and technology evolve.

The initial target of LPWA was applications and services that are not sensitive to delay or data loss. To be widely deployed, LPWA networks must be able to satisfy the requirements of a wide range of services. Examples of services that currently do not have sufficient support of LPWA include video surveillance applications, which require high data rates. Although RPMA and NB-IoT, with their relatively high data rates and large payload length (10 kB \cite{Raza2017} and 317 B \cite{Hoglund2017}, respectively) can be tentative solutions for these applications, they must be permitted to communicate at higher data rates and larger payloads (depending on the encoding rate). This, however, is not possible with convolutional codes in devices that replaced turbo codes for the sake of simple decoding \cite{Hoglund2017}.

Furthermore, because LPWA networks are often capable of supporting low data rates and small frame sizes, the data need to be compressed robustly. It has been proposed to take advantage of compression techniques at the application layer such as CoIP for this purpose \cite{Thubert2017}. LPWA technologies communicate over low data rates that limit the size of frames. This shows the inefficiency of classical compression techniques for these technologies and the need for more efficient techniques. 

To address the requirements of different IoT applications, LoRaWAN offers a number of classes and adapts the data rate using spreading factors. A congestion classifier was proposed by \cite{KIM2017} for data rate control. Although most LPWA technologies support DL communications, some (e.g., SigFox and Telensa) are not effective for use-cases that require communication with end devices due to low data rates on the DL.

LPWA technologies deploy modulation (such as UNB) and spread-spectrum techniques (such as CSS and DSSS) that are resilient to interference. The analysis by \cite{Lieske2016} showed a low error probability for coded time-hopping spread spectrum systems, even under strong interference. LoRa modulation exhibits good resistance to interference \cite{Augustin2016}. However, interference from signals using the same spreading sequence drops the coverage probability exponentially as the number of devices increases	\cite{Georgiou2017}. Furthermore, the popularity of LoRa has resulted in the deployment of multiple networks in close proximity, causing high interference. To ensure the reception of the message, SigFox allows nodes to send the same message up to three times. It has been observed that the use of multiple base stations or directional antennas improves performance in an environment characterized by interference \cite{Voigt2017}. Turbo-FSK is a new UL scheme for LPWA networks. The transmitter performs FSK modulation, and the receiver turbo-decodes the FSK waveforms. M-ary orthogonal FSK modulation is associated with an iterative receiver (the turbo principle) in \cite{Roth2017}. An extrinsic information transfer (EXIT) chart analysis \cite{Brink2001} was used to optimize the parameters and find the best values. The scheme was reported to be energy efficient for different packet sizes when compared with the IEEE 802.15.4k standard \cite{6581828}, the LoRa physical layer \cite{seller2014low} and the serial concatenation of turbo codes (TC) \cite{3GPP2015}. The work extended the initial Turbo-FSK analysis presented in \cite{Roth2015}. Whereas LPWA are required to provide higher data rates and larger payloads to support real-time applications, these come at the expense of increasing the encoding rate and eventually less robust radio links.

Whereas positioning can bring benefits to IoT applications, it is vital in some use-cases. Health condition and environmental monitoring, smart transportation and cities are among use-cases that require the nodes to be able to determine their physical position or logical location. Currently, NB-IoT and LoRaWAN allow the nodes to be located based on received signals and time \cite{Raza2017} \cite{Hoglund2017}. Basic positioning of Release 13 NB-IoT is based on cell identity (CID), and advanced positioning of Release 14 is supported through the observed time difference of arrival (OTDOA) technique \cite{Hoglund2017}. A similar approach with an ultra-high resolution time-stamp is exploited by LoRaWAN; however, its inaccuracy in indoor environments has been argued \cite{Zafari2017}. Mechanisms such as the combination of LoRa and GPS were proposed for more accurate positioning \cite{linklabsLocalization}. This articulates the need for new techniques that can provide precise positioning services to IoT applications.

There are cases where a message needs to be sent to a number of nodes simultaneously, for example, when the firmwares of a large number of end devices are upgraded or street lights are turned on or off. NB-IoT provides multimedia broadcast multicast services (MBMS) using single-cell point-to-multipoint (SC-PtM) data \cite{Hoglund2017}. RPMA and LoRaWAN also support multicast \cite{Ingenu} \cite{Sornin2017}. However, LPWA systems must address issues related to multicast communications such as security and data flooding.

A new provisioning class was recently added to Release 14 NB-IoT to reduce the power consumption of end devices to a maximum of 14 dBm compared to 20 dBm in Release 13 \cite{Hoglund2017}. LoRa was combined with wake-up receivers in an architecture to improve both power consumption and latency \cite{Aoudia2017}. The combination of optimal-hop routing and the transmission configurations algorithm yielded a longer network lifetime for multi-hop communication compared with single-hop communication due to the reduction in the energy consumption of far-end devices in an LPWA network with up to thousands of end devices \cite{BARRACHINAMUNOZ2017153}. Cooperative relaying, signaling, radio resource allocation, and random access schemes are identified as main areas of energy efficiency improvement in \cite{Song2016}. More energy-efficient algorithms are particularly vital for LPWA technologies whose inherent energy efficiency is not high such as RPMA. 
	
Scalability will most likely be a serious issue of LPWA technologies since the number of devices are exponentially increasing. Although LoRa allows transmitters a range of communication options, such as center frequency, SF, bandwidth, and coding rates, it limits the number of supported transmitters. Studies have shown that the number of end devices has a greater impact on LoRa scalability than spectrum restriction \cite{Georgiou2017}. According to \cite{Bor2016} and \cite{lorasim2016}, LoRa can scale well only if it is supported by dynamic transmission parameter selection and/or multiple sinks; otherwise, up to 120 nodes can be handled in 3.8 ha. \\

\section{Conclusion}
\label{Sec:Conclusion}
LPWA technologies have been thoroughly explored in this survey. Unlike other survey papers in the field, our paper argues for the need for integration among diverse LPWA technologies and suggests the most effective solutions for different IoT use-cases. The contribution of these technologies to the market and business is discussed. To provide researchers with the latest in the field, recent research efforts to study or improve LPWA operation have been compared and classified. Although LPWA networks have attractive features for many IoT applications, there are still challenges that need to be addressed. Recommendations on how these issues can be addressed are presented. 

\section*{Acknowledgment}

The authors would like to thank the two anonymous reviewers for their constructive comments and suggestions which have improved the quality of this paper. We would also like to thank Dr. Joseph Ray Carroll for his invaluable comments.

\appendix
\newpage
\section*{ACRONYMS}

\begin{table}[!h]
	\label{tab:acronyms}

	\begin{tabular}{ll}

		6LoWPAN & IPv6-based LoWPAN \\
		AAA & authentication, authorization and accounting \\
		BLE & Bluetooth low energy \\
		BPSK & binary phase shift keying \\
		BR/EDR & Bluetooth basic rate/enhanced data rate \\
		CID & cell identity \\
		CoAP & constrained application protocol \\
		CRC & cyclic redundancy check \\
		CSMA/CA & carrier-sense multiple access with collision avoidance \\
		CSS & chirp spread spectrum \\
		D2D &  device-to-device \\
		DL & downlink \\
		DL MU-MIMO & downlink multi-user MIMO \\
		DRESG & distance-ring exponential stations generator \\
		DSSS & direct sequence spread spectrum \\
		EAP & extensible authentication	protocol \\
		EAP-AKA & EAP-authentication and key agreement\\
		EAP-AKA$'$ & Revision of EAP-AKA \\
		EAP-NOOB & EAP-nimble out-of-band authentication \\
		EAP-PSK & EAP-pre-shared key \\
		EAP-SIM & EAP-subscriber identity modules \\
		EAP-TSL & EAP-transport layer security \\
		EPC & evolved packet core \\
		EXIT & extrinsic information transfer \\
		FEC & forward error correction \\
		FHSS & frequency-hopping spread spectrum \\
		FSK & frequency-shift keying \\
		GFSK & Gaussian frequency shift keying \\
		GPRS & general packet radio service \\
		GSM & global system for mobile communications \\
		HARQ & hybrid automatic repeat request \\
		IoT & Internet of Things \\
		IP & Internet protocol \\
		ISM & industrial, scientific and medical \\
		L2CAP & logical link control and adaptation protocol \\
		LISP & locator identity separation protocol \\
		LMP & link manager protocol \\
		LMSC & LAN/MAN standards committee \\
		LoRa & long-range \\
		LoRaWAN & long-range wide area network \\
		LoWPAN & low-power wireless personal area network \\
		LPWA & low power wide area \\
		LTE & long-term evolution \\
		LTN & low throughput network \\
		M2M & machine-to-machine \\
		MAC & medium access control \\
		MBMS & multimedia broadcast multicast services \\
		MCS & modulation and coding schemes \\
		MIMO & multiple-input multiple-output \\
		NB-IoT & narrow band-IoT \\
		NEMO & network mobility \\
		OFDM & orthogonal frequency division multiplexing \\
		OSSS & orthogonal sequence spread spectrum \\
		OTDOA & observed time difference of arrival \\
		PHY & physical \\
		QoS & quality of service \\
		RF & radio frequency \\
		RPMA & random phase multiple access \\
		SC-PtM & single-cell point-to-multipoint \\
		SDO & standard developing organization \\
		SF & spreading factor \\
		SIG & special interest group \\
		SINR & signal-to-interference-plus-noise ratio \\
		SOLACE & smart object lifecycle architecture for constrained environments \\
		TC & turbo code \\
		UL & uplink \\
		UNB & ultra narrow band \\
		WAN & wide area network \\
		WBAN & wireless body area network \\
		WLAN & wireless local area network \\
		WSN & wireless sensor network \\

	\end{tabular}
\end{table}

\EOD
\bibliographystyle{IEEEtran}


\end{document}